\documentclass[journal]{IEEEtran}
\usepackage[bottom]{footmisc}
\usepackage{amsmath,amssymb,amsfonts}
\usepackage{commath}
\usepackage{setspace}
\usepackage{cite}
\usepackage{graphicx}
\usepackage[center]{caption}
\usepackage{wrapfig}
\usepackage{hyperref}
\usepackage{pgfplots}
\pgfplotsset{width=10cm,compat=1.9}
\usepackage{tikz}
\usetikzlibrary{shapes.geometric}
\usepackage[ruled,vlined,linesnumbered]{algorithm2e}
\usepackage{textcomp}
\usepackage{xcolor}
\def\BibTeX{{\rm B\kern-.05em{\sc i\kern-.025em b}\kern-.08em
    T\kern-.1667em\lower.7ex\hbox{E}\kern-.125emX}}
\definecolor{light-blue}{RGB}{153, 204, 255}
\definecolor{eddA}{rgb}{0.82, 0.1, 0.26}
\definecolor{eddste}{rgb}{0.6, 0.4, 0.8}
\definecolor{ip}{rgb}{0.55, 0.71, 0.0}
\definecolor{greedy}{rgb}{0.0, 0.18, 0.39}
\definecolor{random}{rgb}{0.28, 0.02, 0.03}

\usepackage{scalerel}
\usepackage{tikz}
\usetikzlibrary{svg.path}

\definecolor{orcidlogocol}{HTML}{A6CE39}
\tikzset{
  orcidlogo/.pic={
    \fill[orcidlogocol] svg{M256,128c0,70.7-57.3,128-128,128C57.3,256,0,198.7,0,128C0,57.3,57.3,0,128,0C198.7,0,256,57.3,256,128z};
    \fill[white] svg{M86.3,186.2H70.9V79.1h15.4v48.4V186.2z}
                 svg{M108.9,79.1h41.6c39.6,0,57,28.3,57,53.6c0,27.5-21.5,53.6-56.8,53.6h-41.8V79.1z M124.3,172.4h24.5c34.9,0,42.9-26.5,42.9-39.7c0-21.5-13.7-39.7-43.7-39.7h-23.7V172.4z}
                 svg{M88.7,56.8c0,5.5-4.5,10.1-10.1,10.1c-5.6,0-10.1-4.6-10.1-10.1c0-5.6,4.5-10.1,10.1-10.1C84.2,46.7,88.7,51.3,88.7,56.8z};
  }
}
\newcommand\orcidicon[1]{\href{https://orcid.org/#1}{\mbox{\scalerel*{
\begin{tikzpicture}[yscale=-1,transform shape]
\pic{orcidlogo};
\end{tikzpicture}
}{|}}}}
\ifCLASSINFOpdf
\else
\fi
\begin{document}
\graphicspath{ {./Images/} }
\setstretch{0.9}
%

\title{EDD-NSTE: Edge Data Distribution as a Network Steiner Tree Estimation in Edge Computing}

%
%

\author{Ravi Shankar\textsuperscript{ \orcidicon{0000-0003-4079-4979}}, Aryabartta Sahu\textsuperscript{ \orcidicon{0000-0002-5453-5022  }}~\IEEEmembership{IEEE Senior Member}, \ \\ Deptt. of CSE, IIT Guwahati, Assam, India \ \\ 
email:\{asahu,ravi170101053\}@iitg.ac.in}

%
%

\markboth{IEEE INTERNET OF THINGS}%
{Shell \MakeLowercase{\textit{et al.}}: EDD-NSTE:  Edge Data Distribution as a Network Steiner Tree Estimation in Edge Computing}
%



\maketitle

\begin{abstract}
Edge computing is a distributed computing paradigm that brings computation and data storage closer to the user’s geographical location to improve response times and save bandwidth. It also helps to power a variety of applications requiring low latency. These application data hosted on the cloud needs to be transferred to the respective edge servers in a specific area to help provide low latency app-functionalities to the users of that area. Meanwhile, these arbitrary heavy data transactions from the cloud to the edge servers result in high cost and time penalties. Thus, we need an application data distribution strategy that minimizes these penalties within the app-vendors’ specific latency constraint. \ \\

In this work, we provide a refined formulation of an optimal approach to solve this Edge Data Distribution (EDD) problem using Integer Programming (IP) technique. Due to the time complexity limitation of the IP approach, we suggest an O(k) approximation algorithm  based on network Steiner tree estimation (EDD-NSTE) for estimating solutions to dense large-scale EDD problems. Integer Programming and EDD-NSTE are evaluated on a standard real-world EUA data set and the result demonstrates that EDD-NSTE significantly outperform with a performance margin of 86.67\% over the other three representative approaches and the start of art approach.

\end{abstract}


%
\IEEEpeerreviewmaketitle

\section{Introduction}
\IEEEPARstart{C}{loud} Computing has been dominating the IT discussions for the last two decades, particularly since Amazon popularized the term in 2006 with the release of its Elastic Compute Cloud (Amazon EC2\footnote{Amazon Elastic Compute Cloud (Amazon EC2) is a web service that provides secure, resizable compute capacity in the cloud. It is designed to make web-scale cloud computing easier for developers.}) \cite{IEEE-CC-ACM,amazonec2}. In the simplest form, cloud computing is the practice of using a network of remote servers hosted on the internet to store, manage, and process data, rather than a local server or a personal computer. This facilitates the users to take advantage of the shared data center infrastructure and economy of scale to reduce costs. However, latency influenced by the number of router hops, packet delays introduced by virtualization in the network, or the server placement within a data center, has always been a key issue for cloud migration. Other conventional network paradigms provided by cloud computing, which includes content-centric network, content-delivery network, and information-centric network also cannot handle the huge increase in the latency of the network and the congestion caused by the resources at the edge of the cloud \cite{rfc8793,Pathan2008}. 

This is where edge computing kicks in. Edge Computing is essentially the process of decentralizing the computer services with the help of edge servers deployed at the base stations or access points that are geographically close to the users \cite{IEEEexample:davisEC,IEEE-EC}. This can significantly reduce the latency as it can drastically reduce the volume of data moved and the distance it travels. These computing and storage resources can then be easily hired by any mobile or IoT app-vendor for hosting their applications. The users can then get rid of the computational burden and energy overload from their resource-limited end-devices to the edge servers located nearby. Thus, from an app vendors' perspective, caching data on the edge servers can reduce both the latency of their users to fetch the data and the volume of their application data transmitted between the cloud and its users, thereby, reducing the transmission cost. But then a new challenge arises here as to how to distribute the application data onto the edge servers to minimize the cost incurred for the transactions of data between cloud-to-edge (C2E) and edge-to-edge (E2E) servers.
\begin{figure}[tb!]
\centering
\includegraphics[width=0.4\textwidth]{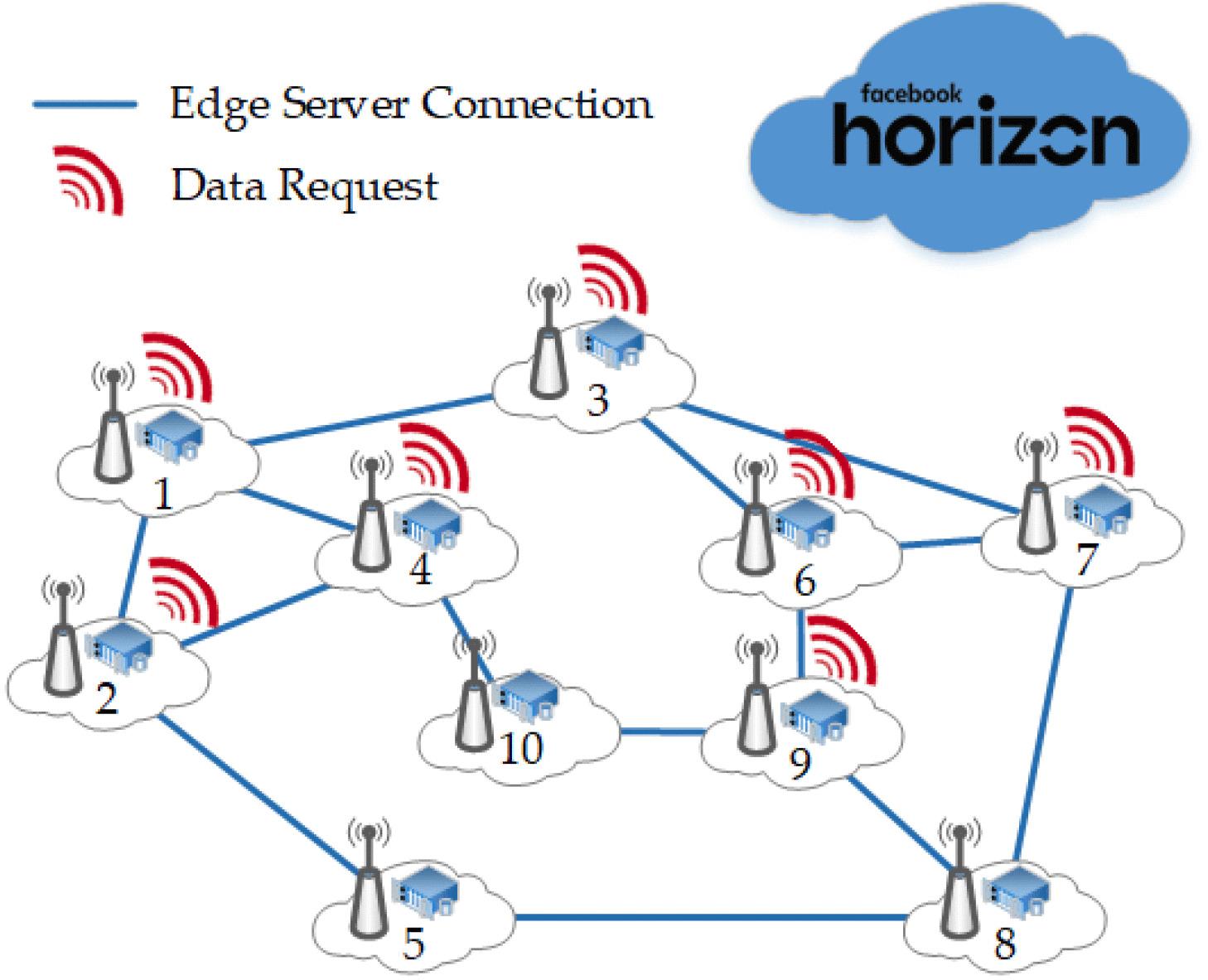}
\caption{An industry example} \label{fig:fig1}
\end{figure}

Facebook Horizon \cite{FH} is an example of industrial application that can benefit considerably from caching their data onto the edge servers. Facebook users using Oculus headsets can access Virtual Reality (VR) videos and VR games on Facebook Horizon. VR users and their applications are highly latency-sensitive. Thus, caching the most trendy and popular VR videos and VR games onto the edge servers will allow the users near the geographical locations of those edge servers to access the application data with minimum latency which in turn will increase the VR experience, sensitivity, and performance. \autoref{fig:fig1} displays a simple graphical example of this idea, where it is required to distribute the application data to edge servers to decrease the data traffic and request-response time between the Facebook Horizon cloud servers and its users in those specific areas. Therefore, the edge data distribution is a must, and at the same time, a cost-effective method to distribute the application data needs to be incorporated as otherwise the cost-ineffective or random app data distribution strategy can cost Facebook Horizon significantly. The cost-effective data distribution strategy should also consider a constraint on the time taken to distribute the application data to each of the edge servers, i.e., the Facebook Horizon latency limit.

The contributions of this research are as follows:
\begin{itemize}
    \item We provide a refined formulation of an optimal approach to solve the Edge Data Distribution problem using the Integer Programming (IP) technique as compared to formulation given in  Xia et al.  \cite{IEEEexample:papermy}.
    \item Since the IP approach has an exponential time complexity, thus, we developed an O(k) approximation algorithm named EDD-NSTE using network Steiner tree estimation (NSTE) for estimating solutions to dense large-scale EDD problems.
    \item The IP and EDD-NSTE solution approaches are then evaluated using standard real-world EUA dataset and the results conclude that EDD-NSTE almost comparable to IP and outperformed the other three approaches in comparison namely \textemdash \hspace{3pt}Greedy, Random, and state-of-the-art EDD-A approach \cite{IEEEexample:papermy}.
\end{itemize}

This paper is arranged in as follows: Section II reviews the previous and related works of this research. Section III describe the edge data distribution problem. In Section IV we discuss the two solution strategies namely, with refined formulation using Integer Programming (IP) and proposed solution EDD-NSTE. Section V evaluates these algorithms along with other representative solution approaches and the state of art approach on the standard real-world EUA dataset, and finally we discuss the concluding remark of our work  in Section VI. 

\section{Related Work}
\label{Sec:RW}

Cloud Computing is the practice of using a network of remote servers hosted on the internet to store, manage, and process data rather than a local server or a personal computer. Cloud-based storage makes it possible to save files to a remote database and retrieve them at user request. The scheduling of these requests is a crucial problem in cloud computing and, over the years, researchers have proposed various scheduling algorithms. In a recent research Zhao et al. \cite{IEEEexample:paperpiyush}, the authors suggested a state-of-the-art scheduler, which not only takes into consideration the load balance but also the application properties for request scheduling.

\setlength{\arrayrulewidth}{1pt}
\begin{table}[tb!]
\caption{Summary of Common Notations} \label{tab:table1}
\setlength\tabcolsep{3pt}
\centering
\smallskip 
\begin{tabular*}{\columnwidth}{@{\extracolsep{\fill}}ll}
\hline
  \textbf{Notation} & \textbf{Description} \\ 
\hline
  $G$ & graph representing a particular region \\
  $N$ & number of edge servers in a particular region\\ 
  $V$ & set of all the edge servers in a particular region\\
  $E$ & set of all edges in the graph, i.e, high-speed links \\
  $R$ & set of destination edge servers in the graph $G$\\
  $u$, $v$ & edge servers \\
  $e\textsubscript{(u, v)}$ & link/edge between edge server nodes $u$ and $v$\\
  $c$ & cloud server \\ 
  $S$ & set of binary variables indicating initial transit\\
      & edge server\\
  $H$ & set of binary variables indicating edge servers\\
      & visited during EDD process\\
  $T$ & set of binary variables indicating the data\\
      & distribution path\\
  $\gamma$ & ratio of cost of C2E to cost of 1-Hop E2E\\
  $\rho$ & destination edge server density \\
  $\delta$ & edge density \\
  $d\textsubscript{limit}$ & vendors' latency constraint, or hop limit\\
  $D\textsubscript{limit}$ & depth limit, $D\textsubscript{limit} = d\textsubscript{limit} + 1$\\
  $D\textsubscript{v}$ & depth of edge server $v$, $D\textsubscript{v} = d\textsubscript{v} + 1$\\
  $K$ & depth limit, $K = D\textsubscript{limit}$\\
  $G\textsubscript{DT}$ & directed graph with cloud as the root and edges\\
                        & replaced by the shortest path between two nodes\\
                        & in graph $G$ \\ 
  $\overline{G}$ & metric closure of graph $G$\\
  $\overline{E}$ & set of edges in the metric closure of graph $G$\\
  $\overline{G}\textsubscript{R}$ & subgraph of $\overline{G}$ induced over $R$\\
  $mst(G)$ & minimum spanning tree of $G$\\
  $Triples$ & set of all combination of 3 destination\\
            & edge servers\\
  $G\textsubscript{cv}$ & tree having cloud as the root and edges\\
                        & $e \in G \setminus \{c\}$\\
  $G\textsubscript{opt}$ & optimal solution of the given EDD problem\\
  $Cost(G)$ & cost associated with any graph $G$\\ 
  $v(z)$ & centroid of a particular $Triple$ $z$\\
  $d(e)$ & weight of edge $e$ in the graph\\
\hline
\end{tabular*}
\end{table}

On the other hand, Edge Computing is an extension of cloud computing which distributes computing resources and services to the edge servers of a particular region. The placement of these edge servers is a fundamental issue in edge computing. In Yao et al. \cite{IEEEexample:paperA}, the authors suggested a cost-effective method that uses 0-1 integer programming to help providers of edge infrastructure make correct decisions regarding the edge servers placement. Similarly, Hao Yin et al. \cite{IEEEexample:paperB} suggested a decision support framework based on a flexible server placement, namely Tentacle, which aims to minimize the cost while maximizing the overall system performance.

The cost-effective application user allocation is another fundamental problem in edge computing first studied in \cite{IEEEexample:paperF}. In research \cite{IEEEexample:papersparsh}, the authors formulated it as a bin packing problem and suggested a heuristic approach for approximating the sub-optimal solutions to this problem.\\ 
In the recent researches \cite{IEEEexample:paperC}, \cite{IEEEexample:paperD}, \cite{IEEEExample:paperX}, \cite{IEEEExample:paperY}, \cite{IEEEExample:paperZ} and \cite{IEEEexample:paperE}, researchers have proposed investigative new techniques and approaches for data caching in the edge computing environment. However, even after having an optimal edge server placement, application user allocation, and data caching, we still need to consider the fact that transmitting the application data from the cloud to the edge servers also contributes a considerable amount to the app vendors' expense structure.

Thus, a cost-effective application data distribution strategy is needed to minimize and accurately estimate the app vendors' total cost. A recent research in Xia et al.  \cite{IEEEexample:papermy}, authors have formulated this problem and suggested a heuristic state-of-the-art algorithm to estimate the total cost, including both the C2E and E2E transmissions. This paper we extends the idea, refined the formulation and introduces a solution approach named EDD-NSTE based on network Steiner tree estimation, for estimating the total cost associated with the edge data distribution problem from the app vendors' perspective.

\begin{figure}[tb!]
\begin{center}
\begin{tikzpicture}[node distance={13.5mm}, thick, main/.style = {draw, circle}, square/.style={regular polygon,regular polygon sides=4}]
\tikzstyle{every node}=[font=\small]
\node[square] (6) [fill=light-blue] {\textbf{6}};
\node[square] (5) [fill=light-blue,right of=6] {\textbf{5}};
\node[square] (9) [below of=5, fill=light-blue] {\textbf{9}};
\node[square] (3) [fill=light-blue, below of=6] {\textbf{3}};
\node[square] (8) [fill=light-blue, above left of=6] {\textbf{8}};
\node[main] (7) [right of=8] {7};
\node[square] (2) [fill=light-blue, below left of=3] {\textbf{2}};
\node[square] (4) [fill=light-blue, below right of=9] {\textbf{4}};
\node[main] (10) [above right of=5] {10};
\node[main] (1) [above of=4] {1};

\draw[-] (6) -- (5);
\draw[-] (3) -- (5);
\draw[-] (3) -- (9);
\draw[-] (3) -- (2);
\draw[-] (8) -- (2);
\draw[-] (8) -- (7);
\draw[-] (8) -- (7);
\draw[-] (7) -- (10);
\draw[-] (5) -- (9);
\draw[-] (2) -- (4);
\draw[-] (4) -- (1);
\draw[-] (1) -- (10);
\draw[-] (10) -- (5);
\draw[-] (8) -- (6);
\draw[-] (9) -- (4);
\end{tikzpicture}
\caption{EDD scenario with 10 edge servers} \label{fig:fig2}
\end{center}
\end{figure}

\setlength{\textfloatsep}{5pt}

\section{Problem Formulation}
\label{sec:PF}

In edge computing, adjacent edge servers at different stations can also communicate, share information as well as resources between them. Thus, edge servers of a specific geo-location can be combined to form a network, which can then be formulated as a graph. In this paper, we consider $N$ edge servers of a specific area and model it as a graph $G$. Each node $v$ of the graph $G$ represents an edge-server $v$, and each edge of the graph connecting any two nodes $u$ and $v$, i.e., $e\textsubscript{(u, v)}$ represents the link between these two edge-servers. Thus, we use $G(V, E)$ to represent the graph, where $V$ is the set of edge servers in the graph and $E$ is the set of links. Let $R$ denote the set of destination edge servers in the graph $G$, i.e., the edge servers which must be sent data directly or indirectly from the cloud servers within the vendor's specific latency constraint. 

For example, in \autoref{fig:fig2}, $N$ = 10, which means we have 10 edge servers i.e., $V$ = \{1, 2, 3, 4, 5, 6, 7, 8, 9, 10\}. The set of edges $E$ = \{$e\textsubscript{(2, 3)}$, $e\textsubscript{(6, 8)}$, $e\textsubscript{(8, 7)}$, $\cdots$, $e\textsubscript{(2, 4)}$\}. Similarly, the set of destination edge servers $R$ = \{2, 3, 4, 5, 6, 8, 9\} shown as square box in the \autoref{fig:fig2}. Common notations used throughout the text in the paper are given in \autoref{tab:table1}.

\begin{figure*}[!tb]
\centering
\begin{minipage}[b]{0.48\textwidth}
\centering
    \begin{tikzpicture}[node distance={16mm}, thick, main/.style = {draw, circle},new/.style = {draw, circle, line width=1.5pt}, square/.style={regular polygon,regular polygon sides=4}]
    \tikzstyle{every node}=[font=\small]
    \node[square] (1) [fill=light-blue] {\textbf{1}};
    \node[new] (2) [right of=1] {2};
    \node[new] (3) [right of=2] {3};
    \node[new] (4) [below of=2] {4};
    \node[new] (5) [below of=1] {5};
    \node[new] (6) [below of=5] {6};
    \node[main] (7) [right of=6] {7};
    \node[new] (8) [below of=3] {8};
    \node[main] (9) [below of=8] {9};
    
    \draw[line width = 2pt, ->] (1) -- (2); 
    \draw[line width = 2pt, ->] (2) -- (3);
    \draw[line width = 2pt, ->] (1) -- (5);
    \draw[line width = 2pt, ->] (5) -- (6);
    \draw[-] (6) -- (7);
    \draw[line width = 2pt, ->] (1) -- (4);
    \draw[line width = 2pt, ->] (4) -- (6);
    \draw[-] (7) -- (9);
    \draw[-] (3) -- (8);
    \draw[-] (8) -- (9);
    \draw[line width = 2pt, ->] (4) -- (8);
    \end{tikzpicture}
    \caption{EDD example to demonstrate d\textsubscript{limit}} \label{fig:fig3}
\end{minipage}
\begin{minipage}[b]{0.48\textwidth}
\centering
    \begin{tikzpicture}[node distance={13mm}, thick, main/.style = {draw, circle}, square/.style={regular polygon,regular polygon sides=4}]
    \tikzstyle{every node}=[font=\small]
    \node[square] (6) [fill=light-blue] {\textbf{6}};
    \node[square] (5) [fill=light-blue,right of=6] {\textbf{5}};
    \node[square] (9) [below of=5, fill=light-blue] {\textbf{9}};
    \node[square] (3) [fill=light-blue, below of=6] {\textbf{3}};
    \node[square] (8) [fill=light-blue, above left of=6] {\textbf{8}};
    \node[main] (7) [right of=8] {7};
    \node[square] (2) [fill=light-blue, below left of=3] {\textbf{2}};
    \node[square] (4) [fill=light-blue, below right of=9] {\textbf{4}};
    \node[main] (10) [above right of=5] {10};
    \node[main] (1) [above of=4] {1};
    
    \draw[-] (6) -- (5);
    \draw[-] (3) -- (5);
    \draw[line width = 2pt, <-] (3) -- (9);
    \draw[-] (3) -- (2);
    \draw[line width = 2pt, ->] (8) -- (2);
    \draw[-] (8) -- (7);
    \draw[-] (8) -- (7);
    \draw[-] (7) -- (10);
    \draw[line width = 2pt, <-] (5) -- (9);
    \draw[-] (2) -- (4);
    \draw[-] (4) -- (1);
    \draw[-] (1) -- (10);
    \draw[-] (10) -- (5);
    \draw[line width = 2pt, ->] (8) -- (6);
    \draw[line width = 2pt, ->] (9) -- (4);
    \end{tikzpicture}
    \caption{Optimal solution using integer programming} \label{fig:fig4}
\end{minipage}
\end{figure*}

Lead cloud service providers like Google, Amazon, or Microsoft charge differently for data transactions. For e.g., hosting applications in Microsoft Azure involves careful consideration of the costs generated by three main components i.e., computing, storage, and network. Similarly, for any edge infrastructure provider, the pricing models usually are different for cloud-to-edge and edge-to-edge transmissions. Thus, we need to define a relationship between these two transmissions, so that they can be incorporated easily into the generic optimization objective model. We define $\gamma$, to specify the ratio between the C2E and E2E transmission costs. For e.g., $\gamma = 20$, means the $Cost\textsubscript{C2E}$ is 20 times more than $Cost\textsubscript{E2E}$. The cloud-to-edge transmissions are much more expensive than the edge-to-edge transmissions mainly because the graph nodes are connected via high-speed transmission links and are also not that distant apart from each other. Since the transmission latency is measured by the number of hops it takes to reach the destination node from the source in graph $G$, the C2E cost can then be easily transformed to $\gamma$ times of the 1-hop E2E transmission cost in the generic model.

Now there are two possible scenarios for data transmissions in EDD, firstly C2E, where the data is sent from the cloud to some of the edge servers referred to as the  \textit{"initial transit edge servers"} and secondly E2E, where the data is transmitted between any two edge servers\footnote{There is no necessity for initial transit edge servers to be a destination edge server and a destination edge server can also be an initial transit edge server as it may transfer data further to other servers.}. Thus, an EDD strategy comprises of two parts, a C2E and an E2E strategy.
\begin{itemize}
    \item In the C2E strategy, the set $S$ = \{$s$\textsubscript{1}, $s$\textsubscript{2}, $\cdots$, $s$\textsubscript{N}\}, where $s\textsubscript{v}$ (1 $\leq$ $v$ $\leq$ $N$) denotes the set of Boolean array representation of the initial transit edge servers, which receives data directly from the cloud.
\begin{equation} \label{eq:constraint1}
    s\textsubscript{v} = 
    \begin{cases}
    1, & \text{if \textit{v} is an initial transit edge server}\\
    0, & \text{if \textit{v} is not an initial transit edge server}
    \end{cases}
\end{equation}
\item  Similarly, In the E2E strategy, the set $T$ = \{$\tau$\textsubscript{(1, 1)}, $\tau$\textsubscript{(1, 2)}, $\cdots$, $\tau$\textsubscript{(N, N)}\}, where $\tau$\textsubscript{\textit{(u, v)}} (\textit{u, v} $\in$ \textit{V}) denotes whether the data is transmitted through edge \textit{e\textsubscript{(u, v)}} in \textit{G}.
\begin{equation} \label{eq:constraint2}
    \tau\textsubscript{\textit{(u, v)}} = 
    \begin{cases}
    1, & \text{if data is transmitted through {e\textsubscript{(u, v)}}}\\
    0, & \text{if data is not transmitted through {e\textsubscript{(u, v)}}}
    \end{cases}
\end{equation}
Since a reasonable EDD solution must link each destination edge server \textit{v} $\in$ \textit{R} to an initial transit edge server in \textit{S} through edges in \textit{E}, thus,
\begin{equation} \label{eq:constraint3}
    \textit{Connected(S, T, u, v)} = true, \hspace{3pt} \forall \textit{v} \in \textit{R}, \hspace{3pt} \exists \hspace{2pt} u, s\textsubscript{u} = 1
\end{equation}
\end{itemize}

Now, the app-vendor has the liberty to regulate the EDD latency constraint as per the application specifics and requirements. Let $d\textsubscript{limit}$ represent the app vendors EDD time constraint. Thus, for each of the destination edge server $v$, edge-to-edge latency or data transmission latency or path length in hops between $v$ and its parent initial transit edge server in $S$, i.e., $d\textsubscript{v}$ should not exceed this time or depth constraint.
\begin{equation} \label{eq:constraint4}
    0 \leq \textit{d\textsubscript{v}} \leq \textit{d\textsubscript{limit}}, \textit{d\textsubscript{v}} \in Z\textsuperscript{+}, \forall \textit{v} \in \textit{R}
\end{equation}
For example, consider the graph in \autoref{fig:fig3}, the nodes of the graph represent the edge servers given by
$V$ = \{1, 2, 3, 4, 5, 6, 7, 9\}, where number of edge servers $N$ = 9 and the set of high-speed links i.e., $E$ = \{$e$\textsubscript{(1, 2)}, $e$\textsubscript{(1, 3)}, $e$\textsubscript{(1, 4)}, $\cdots$, $e$\textsubscript{(7, 9)}\}. Now, let the EDD time constraint as per the vendor's application requirements be $d\textsubscript{limit}$ = 2, which means that it should take less than or equal to two hops for a destination edge server to receive the data from any initial transit edge server. In \autoref{fig:fig3}, if node 1 is the only node selected as the initial transit edge server, then one possible E2E strategy is to select edges \{$e$\textsubscript{(1, 2)}, $e$\textsubscript{(1, 4)}, $e$\textsubscript{(1, 5)}, $e$\textsubscript{(5, 6)}, $e$\textsubscript{(2, 3)}, $e$\textsubscript{(4, 6)}, $e$\textsubscript{(4, 8)}\}. This means that in the set T\textsubscript{E2E} the value of  $\tau$\textsubscript{(1, 2)} =  $\tau$\textsubscript{(1, 4)} = $\tau$\textsubscript{(1, 5)} = $\tau$\textsubscript{(2, 3)} = $\tau$\textsubscript{(4, 6)} = $\tau$\textsubscript{(5, 6)} = $\tau$\textsubscript{(4, 8)} = 1. Accordingly, we can also obtain the time constraint limit of each of these destination edge servers, i.e, $d$\textsubscript{1} = 0, $d$\textsubscript{2} = $d$\textsubscript{4} = $d$\textsubscript{5} = 1, and $d$\textsubscript{6} = $d$\textsubscript{8} = $d$\textsubscript{3} = 2.

Now, given an EDD latency constraint of $d\textsubscript{limit}$, the app vendor’s aim is to minimize the EDD cost, which consists of the part incurred by the C2E transmissions and the E2E transmissions.

Thus, the Edge Data Distribution (EDD) problem can be formulated as \textemdash
\begin{equation}
    minimize \left\{ Cost\textsubscript{C2E}(S) + Cost\textsubscript{E2E}(T) \right\}
\end{equation}
while fulfilling the EDD latency constraint in \autoref{eq:constraint4}.

As we can see, the Edge Data Distribution (EDD) is a constrained optimization problem based on a well-known NP-hard problem in graph theory known as the Steiner Tree \cite{IEEEexample:papermy}.

\section{Solution Strategy}
\label{sec:SS}
\hypertarget{SolSgy}{}

{In} this section, we first present a refined formulation of the optimization approach given in Xia et al.  \cite{IEEEexample:papermy} where the Edge Data Distribution (EDD) is formulated as a constrained integer programming problem which can then be solved by using any simple IP solver\footnote{IP solvers are the frameworks or tools used to solve integer programming problems. For E.g., \href{https://www.gnu.org/software/glpk/}{GLPK}, \href{http://lpsolve.sourceforge.net/5.5/}{LP Solve}, \href{https://github.com/coin-or/Clp}{CLP}, \href{https://github.com/coin-or/Cbc}{CBC}, \href{https://cvxopt.org/}{CVXOPT}, \href{https://docs.scipy.org/doc/scipy/reference/optimize.html}{SciPy}, \href{https://www.gurobi.com/}{Gurobi Optimizer}, \href{https://www.ibm.com/analytics/cplex-optimizer}{CPLEX}, \href{https://www.fico.com/en/products/fico-xpress-solver}{XPRESS}, \href{https://www.mosek.com/}{MOSEK}, \href{https://developers.google.com/optimization}{Google OR Tools}, \href{https://pypi.org/project/PuLP/}{PuLP}, etc.}. Since this method takes an exponential amount of time to produce results for large datasets, we designed an $O(k)$ approximation method, named EDD-NSTE that can estimate solutions to the EDD problems in polynomial time. Also, we did the theoretical analysis of the performance of the proposed EDD-NSTE based solution approach.

\subsection{Edge Data Distribution (EDD) as an Integer Programming problem}

The Edge Data Distribution (EDD) can be formulated as a constrained integer programming problem as given in \cite{IEEEexample:papermy}. Our refined formulation of the same is specified as below.

The solution for the EDD should minimize the cost incurred for data transmissions within the app vendor’s latency constraint of $d$\textsubscript{limit}. Thus, to model the EDD problem as a generalized constrained integer programming optimization problem firstly we add the cloud server $c$ into $V$, and then add the edges from cloud $c$ to each edge server in graph $G$. Now, we can formulate the C2E strategy of EDD, $S$ = \{$s$\textsubscript{1}, $s$\textsubscript{2}, $s$\textsubscript{3} $\cdots$ $s$\textsubscript{N}) by selecting edges in graph $G$, such that, T\textsubscript{c} = ($\tau$\textsubscript{(c, 1)}, $\tau$\textsubscript{(c, 2)}, $\cdots$,  $\tau$\textsubscript{(c, N)}\}, where $\tau$\textsubscript{(c, v)} denotes if the data is transferred from cloud server $c$ to the edge server $v$. Here, we combine the C2E and E2E strategy of EDD into one parameter
\begin{equation}
\begin{gathered}
    T = \{\tau\textsubscript{(c, 1)}, \tau\textsubscript{(c, 2)} \cdots  \tau\textsubscript{(c, N)}, \tau\textsubscript{(1, 1)}, \cdots,  \tau\textsubscript{(N, N)}\},\\
    where, \hspace{3pt} \tau\textsubscript{(u, v)} \in (0, 1) \hspace{3pt} (u \in V, v \in V \setminus\{c\})
\end{gathered}
\end{equation}
which indicates whether the edge $e$\textsubscript{(u, v)} is included in $T$. Secondly, we need to update the definition of EDD time constraint as the depth of the edge server in the graph with root as cloud. So, $D$\textsubscript{$v$} represents the depth of the edge server $v$, where $D$\textsubscript{$v$} = $d$\textsubscript{$v$} + 1 and $D$\textsubscript{$c$} = 0, as the cloud is the root of the graph. Thus, we can define the limit as 
\begin{equation}
D\textsubscript{limit} = d\textsubscript{limit} + 1 
\end{equation}
Also, we need to define a parameter for each of the edge servers, 
\begin{equation}
    H\textsubscript{v} = 
    \begin{cases}
    0, & \text{if v is not visited during the EDD process}\\
    1, & \text{if v is visited during the EDD process}
    \end{cases}
\end{equation}
here, $H\textsubscript{v} = 1, \forall v \in R$ make sure that it include the destination edge servers into our solution. Also, $H$\textsubscript{$c$} = 1, as cloud  always be a part of the solution. Now, the constrained integer programming optimization problem can be formally expressed as follows, 
\begin{equation}\label{eq:eqMin}
    min \left( \mathop{\gamma\sum}_{v \hspace{2pt} \in \hspace{2pt} V \setminus \{c\}}\hspace{-5pt}\tau\textsubscript{(c, v)} \hspace{4pt} + \mathop{\sum\sum}_{u, v \hspace{2pt} \in \hspace{2pt} V \setminus \{c\}}\hspace{-5pt}\tau\textsubscript{(u, v)} \right)
\end{equation}
\begin{flushleft}
subjected to constraints,
\end{flushleft}
\begin{equation}
    H\textsubscript{v} = 1, \forall v \in R \cup \{c\}
\end{equation}
\begin{equation}\label{eq:const1}
    \tau\textsubscript{(u, v)} \leq H\textsubscript{u}\cdot H\textsubscript{v}, \forall u \in V, v \in V \setminus \{c\}
\end{equation}
\begin{equation}\label{eq:const2}
    \sum \tau\textsubscript{(u, v)} = H\textsubscript{v}, \forall u \in V,  v \in V \setminus\{c\}
\end{equation}
\begin{equation}\label{eq:const3}
    \mathop{\sum_{v \in V \setminus \{c\}}\hspace{-9pt}\tau\textsubscript{(c, v)}} \geq 1
\end{equation}
\begin{equation}\label{eq:const4}
    \tau\textsubscript{(u, v)}, H\textsubscript{v} \in (0, 1)
\end{equation}
\begin{equation}\label{eq:const5}
    \tau\textsubscript{(c, v)} = 1 \hspace{3pt} \forall v \in V\setminus \{c\},D\textsubscript{v} = 1 
\end{equation}
\begin{equation}\label{eq:const6}
    D\textsubscript{c} = 0, D\textsubscript{c} \leq D\textsubscript{v} \leq D\textsubscript{limit}, \forall v \in V \setminus \{c\}
\end{equation}
\begin{equation}\label{eq:const7}
    D\textsubscript{v} - D\textsubscript{u} = 1, \forall u, v \in V, \tau\textsubscript{(u, v)} = 1
\end{equation}

Here, in-order to minimize the total cost incurred by both the C2E and E2E edges given by \autoref{eq:eqMin}, we must satisfy the above constraints. \autoref{eq:const1}  make sure that, if edge $e\textsubscript{(u, v)}$ is considered a part of the solution, or, $\tau\textsubscript{(u, v)} = 1$, then both the nodes $u$ and $v$ must be visited, i.e., $H\textsubscript{v} = 1$ and $H\textsubscript{u} = 1$. \autoref{eq:const2} suggests a constraint that the summation of $\tau\textsubscript{(u, v)}$ of the edges incoming at the particular node $v$, must be equal to the $H\textsubscript{v}$ of that particular node, i.e., whether or not the node $v$ is visited, if edges incoming onto it are considered a part of the solution. \autoref{eq:const3} suggests that the solution must have atleast one cloud server, i.e., summation of $\tau\textsubscript{(c, v)}$ must be greater than or equal to $1$. \autoref{eq:const5} makes sure that the node servers having depth of $1$, must be connected to the cloud server. \autoref{eq:const6} suggests that the depth of any edge server $v$, must be within $[0, d\textsubscript{limit}]$, and \autoref{eq:const7} suggests that for nodes $u$, $v$ connected through edge $e\textsubscript{(u, v)}$ in the EDD solution must satisfy the depth difference of $1$.

Now, this Integer Programming model is a generalized model and thus, for vendor’s specific cost models say \textit{cost(c, v)} and \textit{cost(u, v)} can be simply incorporated into our \autoref{eq:eqMin}, as 
\begin{center}
\begin{gather}
min \left( \mathop{\sum}_{v \hspace{2pt} \in \hspace{2pt} V \setminus \{c\}}\hspace{-9pt}\tau\textsubscript{(c, v)} \cdot cost\textsubscript{(c, v)} \hspace{4pt} + \mathop{\sum\sum}_{u, v \hspace{2pt} \in \hspace{2pt} V \setminus \{c\}}\hspace{-5pt}\tau\textsubscript{(u, v)} \cdot cost\textsubscript{(u, v)} \right)
\end{gather}
\end{center}
Here, as we know $D$\textsubscript{limit} $\geq$ 1, as otherwise the cloud cannot transfer data to any edge server, and for the base case of $D$\textsubscript{limit} = 1, the cost incurred will be $\gamma$R times the cost of 1-hop E2E transmission.

Here in the refined formulation \autoref{eq:const3}, \autoref{eq:const4}, and \autoref{eq:const5} are the new constraints added to the formulation of the integer programming method given in \cite{IEEEexample:papermy}. These constraints narrow the possible solution search-space and also updates the final solution to contain all the edges, i.e., the C2E edges and the E2E edges.

\subsection{Example to understand Integer Programming}

Consider an example in \autoref{fig:fig4} that displays the optimal solution generated, if we solve the graph in \autoref{fig:fig2} using the integer programming approach, with EDD time constraint of \textit{d\textsubscript{limit} = 1}. The C2E strategy specify \textit{node 8} and \textit{node 9} as the initial transit edge servers which receives the data directly from the cloud server and thus, the cost incurred for the cloud to edge data transmission is \textit{Cost\textsubscript{C2E}} = \textit{2}$\gamma$ times of the 1-hop E2E transmission cost. Now, these nodes transmit the data to the other destination edge servers (represented by light blue square nodes) that can be reached within the given time constraint of \textit{d\textsubscript{limit}}. Thus, the cost incurred for the edge to edge data transmission is \textit{Cost\textsubscript{E2E} = 5} times of the 1-hop E2E transmission cost. The integer programming approach selects the edges E = \{e\textsubscript{(c, 8)}, e\textsubscript{(c, 9)}, e\textsubscript{(8, 2)}, e\textsubscript{(8, 6)}, e\textsubscript{(9, 3)}, e\textsubscript{(9, 4)}, e\textsubscript{(9, 5)}\}, for data transmissions and thus, the optimal total cost incurred is the sum of \textit{Cost\textsubscript{C2E}} and \textit{Cost\textsubscript{E2E}}, i.e., \textit{2}$\gamma$ \textit{+ 5} times of the 1-hop E2E transmission cost. 

\subsection{Edge Data Distribution (EDD) as a Network Steiner Tree 
Estimation problem}

\begin{algorithm}[tb!]
    \caption{\textbf{Network Steiner Tree Approximation}}
        F $\gets \overline{G}\textsubscript{R}$;\hspace{3pt}W $\gets$ $\phi$;\hspace{3pt}$Triples$ $\gets$ \{$z$ $\subset$ R :\norm{z}= 3\} \\
        \ForEach{z $\in$ $Triples$}{
            \textbf{find} v which \textbf{minimizes} the $\sum_{s \in z} d(v, s)$ \\
            v(z) $\gets$ v \\
            d(z) $\gets$ $\sum_{s \in z} d(v(z), s)$ \\
        }
        \While{true}{
            M $\gets$ an MST of F; $FindSave(M)$ \\
            \textbf{find} z $\in$ $Triples$ which \textbf{maximizes}, $win$ = $\max\limits_{e \in z}save(e)$ + $\min\limits_{e \in z}save(e)$ - d($z$)\\ 
            \If{win $\leq$ 0}{\textbf{break}}
            F $\gets$ F[z]; insert(W, v(z)) \\
        }
        Find a steiner tree for R $\cup$ W in graph G using MST algorithm
\end{algorithm}
\begin{algorithm}[b]
    \caption{\textbf{To calculate $FindSave(M)$}}
    \If{$M$ $\neq$ empty}{
        $e$ $\gets$ an edge of $M$ with maximum weight;\\ $x$ $\gets$ $d(e)$ \\
        $M$\textsubscript{1}, $M$\textsubscript{2} $\gets$ the subgraphs connected through $e$ \\
        \ForEach{vertex $v$\textsubscript{1} of $M$\textsubscript{1} and $v$\textsubscript{2} of $M$\textsubscript{2}}{
            $save$($\overline{e}$) $\gets$ $x$, where $\overline{e}$ = $e\textsubscript{($v$\textsubscript{1}, $v$\textsubscript{2})}$ \\
        }
        $FindSave$($M$\textsubscript{1}); $FindSave$($M$\textsubscript{2})
    }
\end{algorithm}
\begin{algorithm}[tb!]
\caption{\textbf{EDD-NSTE Algorithm}}
\KwIn{G, G\textsubscript{DT}, D\textsubscript{limit}, depth, cost, visited}
\KwOut{G\textsubscript{final}}
Construct the graph $G\textsubscript{ST}$ from \hyperlink{algo:algo1}{Algorithm 1}. Take the maximum connectivity node in the graph $G\textsubscript{ST}$ and connect it to the cloud to make it a rooted tree. Update the distances between consecutive nodes by the shortest path between them and let this graph be $G\textsubscript{DT}$\\
Initialize depth[c] $\gets$ 0, visited[c] $\gets$ 1, depth[v] $\gets$ $\infty$, visited[v] $\gets$ 0, $\forall$ v $\in$ V, D\textsubscript{limit} $\gets$ d\textsubscript{limit} + 1, Update G\textsubscript{DT} to include the minimum cost path between any two vertices with edges in graph G \\
Update G\textsubscript{DT} to form a directed rooted tree, with cloud as the root \\
Run a BFS to compute the minimum depth of each vertex v $\in$ G\textsubscript{DT} from cloud \\
\ForEach{vertex v, in the path from root to leaf node and parent p in Depth-First Search Algorithm}{
    \If{visited[v] $\neq$ 1}{
        \If{depth[v] $\leq$ D\textsubscript{limit} and v $\in$ R}{
            add edge e\textsubscript{(p, v)} in G\textsubscript{final} \\
            visited[v] $\gets$ 1
        }
        \ElseIf{v $\notin$ R}{
            add edge e\textsubscript{(p, v)} in G\textsubscript{final} \\
            visited[v] $\gets$ 1
        }
        \ElseIf{depth[v] $>$ D\textsubscript{limit} and v $\in$ R}{
            depth[v] $\gets$ 1, visited[v] $\gets$ 1\\
            add edge e\textsubscript{(c, v)} in G\textsubscript{final}\\
            \ForEach{child node u of node v in graph G in DFS order}{
                \If{depth[u] $>$ depth[v] + 1 and depth[v] + 1 $\leq$ D\textsubscript{limit}}{
                    \If{e(v, u) $\notin$ G\textsubscript{DT}}{
                        update edge in G\textsubscript{DT} from v to u\\
                        depth[u] $\gets$ depth[v] + 1
                    }
                }
            }
            run BFS and update depths of the vertices in G\textsubscript{DT}
        }
    }
}
Run a DFS and remove unnecessary edges from graph G\textsubscript{final} \\
Calculate total cost incurred as sum of cost[e], $\forall$ e $\in$ G\textsubscript{final} 
\end{algorithm}

The Edge Data Distribution (EDD) is an NP-Hard \cite{IEEEexample:papermy} problem. Thus, finding an optimal solution for large-scale EDD scenarios is troublesome due to the exponential time complexity of the algorithm. To address this issue, we propose an approach, named Edge Data Distribution as a Network Steiner Tree Estimation (EDD-NSTE), to estimate large-scale EDD solutions. The estimation ratio for EDD-NSTE is $O(k)$, which means that the ratio of the cost by EDD-NSTE solution and that of the optimal solution is $O(k)$ in the worst case, where $k$ is constant.

In our proposed approach, we calculates the estimated cost incurred for a given graph in two steps \textemdash
\begin{itemize}
    \item First of all, we calculate a network Steiner tree approximation \cite{IEEEexample:paperZelikovsky} with destination edge servers $R$, as the set of distinguished vertices, in the given graph $G$.
    \item Then, we use an approach which estimates a rooted minimum Steiner tree, by adding cloud $c$ to the graph, and then slicing and fine-tuning the rooted Steiner tree with the vendor's latency constraint of $D\textsubscript{limit}$.
\end{itemize}

\begin{figure*}[tb!]
\centering
\begin{minipage}[b]{0.48\textwidth}
\centering
    \begin{tikzpicture}[node distance={13.5mm}, thick, main/.style = {draw, circle}, square/.style={regular polygon,regular polygon sides=4}]
    \tikzstyle{every node}=[font=\small]
    \node[square] (6) [fill=light-blue] {\textbf{6}};
    \node[square] (5) [fill=light-blue,right of=6] {\textbf{5}};
    \node[square] (9) [below of=5, fill=light-blue] {\textbf{9}};
    \node[square] (3) [fill=light-blue, below of=6] {\textbf{3}};
    \node[square] (8) [fill=light-blue, above left of=6] {\textbf{8}};
    \node[main] (7) [right of=8] {7};
    \node[square] (2) [fill=light-blue, below left of=3] {\textbf{2}};
    \node[square] (4) [fill=light-blue, below right of=9] {\textbf{4}};
    \node[main] (10) [above right of=5] {10};
    \node[main] (1) [above of=4] {1};
    
    \draw[line width = 2pt, -] (6) -- (5);
    \draw[line width = 2pt, -] (3) -- (5);
    \draw[line width = 2pt, -] (3) -- (9);
    \draw[line width = 2pt, -] (3) -- (2);
    \draw[line width = 2pt, -] (8) -- (2);
    \draw[-] (8) -- (7);
    \draw[-] (8) -- (7);
    \draw[-] (7) -- (10);
    \draw[-] (5) -- (9);
    \draw[line width = 2pt, -] (2) -- (4);
    \draw[-] (4) -- (1);
    \draw[-] (1) -- (10);
    \draw[-] (10) -- (5);
    \draw[-] (8) -- (6);
    \draw[-] (9) -- (4);
    \end{tikzpicture}
    \caption{Steiner Tree estimated by Algorithm 1}
    \vspace{-0.5cm}
    \label{fig:ns1}
\end{minipage}
\begin{minipage}[b]{0.48\textwidth}
\centering
    \begin{tikzpicture}[node distance={13.5mm}, thick, main/.style = {draw, circle}, square/.style={regular polygon,regular polygon sides=4}]
    \tikzstyle{every node}=[font=\small]
    \node[square] (6) [fill=light-blue] {\textbf{6}};
    \node[square] (5) [fill=light-blue,right of=6] {\textbf{5}};
    \node[square] (9) [below of=5, fill=light-blue] {\textbf{9}};
    \node[square] (3) [fill=light-blue, below of=6] {\textbf{3}};
    \node[square] (8) [fill=light-blue, above left of=6] {\textbf{8}};
    \node[main] (7) [right of=8] {7};
    \node[square] (2) [fill=light-blue, below left of=3] {\textbf{2}};
    \node[square] (4) [fill=light-blue, below right of=9] {\textbf{4}};
    \node[main] (10) [above right of=5] {10};
    \node[main] (1) [above of=4] {1};
    
    \draw[line width = 2pt, <-] (6) -- (5);
    \draw[-] (3) -- (5);
    \draw[-] (3) -- (9);
    \draw[line width = 2pt, <-] (3) -- (2);
    \draw[line width = 2pt, <-] (8) -- (2);
    \draw[-] (8) -- (7);
    \draw[-] (8) -- (7);
    \draw[-] (7) -- (10);
    \draw[line width = 2pt, ->] (5) -- (9);
    \draw[line width = 2pt, ->] (2) -- (4);
    \draw[-] (4) -- (1);
    \draw[-] (1) -- (10);
    \draw[-] (10) -- (5);
    \draw[-] (8) -- (6);
    \draw[-] (9) -- (4);
    \end{tikzpicture}
    \caption{Final EDD solution estimated by Algorithm 3}
    \vspace{-0.5cm}
    \label{fig:ns2}
\end{minipage}
\end{figure*}

The proposed approach works as follows: Given a graph $G$, we first calculate the metric closure of the graph, denoted by $\overline{G}$, which is a complete graph of $G$, having edge lengths equal to the shortest distance between the vertices in $G$. $\overline{G}\textsubscript{R}$ denotes the subgraph of $\overline{G}$ induced by a vertex subset $R$ $\subseteq$ $V$, i.e., the subgraph on destination edge servers. We denote the minimum spanning tree (MST) of any graph $G$ as $mst(G)$. Now, given a metrically closed graph $G$, we may contract an edge $e$, i.e., reduce its length to 0, we represent such a resulting graph as $G[e]$. For any triple set of vertices $z$ = ($u$, $v$, $w$), the graph $G[z]$ represent a graph resulting from contraction of two edges i.e., $e$\textsubscript{(u, v)} and $e$\textsubscript{(v, w)}. $Triples$ denotes the set of all combination of 3 destination edge servers, $v(z)$ denotes the centroid of a particular triple, i.e., the vertices other than the destination edge servers whose sum of distance from a triple is minimum and $d(e)$ denotes the weight of the edge $e$. \ \\

The Pseudo code of the proposed approach is given in \hyperlink{algo:algo1}{Algorithm 1}, which loops continuously to find the best possible reduction in the cost of the minimum spanning tree of $F$,  where the initial value of $F = \overline{G}\textsubscript{R}$. The algorithm does so by adding the three edges of $G$ having a common end node, i.e., the centroid, and consecutively removing the max-weight edge from each resulting cycle using the $save$ matrix, from the MST of $F$. This class of algorithms is also known as the Triple Loss Contracting Algorithm \cite{IEEEexample:collectionpaperstudy}  \cite{IEEEexample:paperZelikovsky}. The output of \hyperlink{algo:algo1}{Algorithm 1} is an approximated Steiner tree $G\textsubscript{ST}$, with vertices in R $\cup$ W and edges $e$ $\subset$ $\overline{E}$.

Here, we calculate $save$ matrix for edges in the graph $F$, to estimate the maximum win for a given triple, where $save(\overline{e})$, with $\overline{e} = e\textsubscript{(v\textsubscript{1}, v\textsubscript{2})}$ is the minimum value, such that both ends $v\textsubscript{1}$ and $v\textsubscript{2}$ are in the same component of $F$, given by \autoref{eq:saveEq}. To calculate the $save$ matrix, a pseudo code for a recursive function $FindSave(M)$ is given in \hyperlink{algo:algo2}{Algorithm 2}. The implementation time complexity of this algorithm is $O(\norm{R}\textsuperscript{2})$.\\[-10px]
\begin{equation} \label{eq:saveEq}
    save(\overline{e}) = mst(F) - mst(F[\overline{e}]),\overline{e} = e\textsubscript{($v\textsubscript{1}$, $v\textsubscript{2}$)}, \forall \hspace{2pt} v\textsubscript{1}, v\textsubscript{2} \in F
\end{equation}

Once we get an approximated Steiner tree, it is necessary to add cloud server $c$ to the given graph by adding an edge from cloud to the maximum connectivity node in the approximated Steiner tree. Then we convert  resulting graph into a rooted tree, with cloud as the root. We perform the same by a breadth-first search on $G\textsubscript{ST}$. In this process, we also convert the edges between any two nodes of $G\textsubscript{ST}$ by the shortest path between two nodes in graph $G$, as the Steiner tree $G\textsubscript{ST}$, was approximated over the metric closure of graph $G$. Let this resultant directed graph be $G\textsubscript{DT}$. 

The pseudo code for slicing and fine-tuning the graph to satisfy the vendor's latency constraint is shown in \hyperlink{algo:algo3}{Algorithm 3}. In this approach, we take the directed Steiner tree $G\textsubscript{DT}$ and initialize the $depth[]$ and the $visited[]$ for each node of the graph, with $visited[c] = 1$. It then runs the Breadth-First Search to compute the minimum depths of each node assuming the cloud server as the root. After that, it runs the Depth-First Search and \textemdash
\begin{itemize}
	\item for each unvisited destination edge server with depth less than or equal to $D\textsubscript{limit}$, it adds the corresponding edge into the final solution $G\textsubscript{final}$.
	\item for each unvisited edge server, which is not a destination edge server, it again adds the corresponding edge into the final solution $G\textsubscript{final}$.
	\item for each unvisited destination edge server having a depth greater than $D\textsubscript{limit}$, it connects that edge server directly to the cloud and adds edge from cloud to this edge server in $G\textsubscript{final}$. It then runs a Depth-First Search to update the minimum depths of the remaining unvisited edge servers assuming this edge server as the root.
\end{itemize}
Finally, this algorithm removes the unnecessary edges from the graph $G\textsubscript{final}$ and calculates the cost of the solution as the sum of the cost of edges in $G\textsubscript{final}$.

\subsection{Example to understand EDD-NSTE algorithm}

Let us consider the graph in the \autoref{fig:fig2}, with vendors' latency constraint or hop-limit of $d\textsubscript{limit} = 1$. When we apply \hyperlink{algo:algo1}{Algorithm 1} onto it, it selects edges \{e\textsubscript{(2, 3)}, e\textsubscript{(2, 4)}, e\textsubscript{(2, 8)}, e\textsubscript{(3, 5)}, e\textsubscript{(3, 9)}, e\textsubscript{(5, 6)}\} to construct the estimated Steiner tree given in \autoref{fig:ns1}, using the triple loss contracting mechanism, which in this case is also the optimal Steiner tree possible. Now, as we can see the \textit{node 2} is the node with the max-connectivity in the estimated steiner tree, thus, we add this node to the cloud directly, as given in \autoref{fig:ns2} to make a rooted EDD solution. The \textit{node 3}, \textit{node 4}, and \textit{node 8} can be reached within the 1-hop distance from \textit{node 2}, thus edges \{e\textsubscript{(c, 2)}, e\textsubscript{(2, 3)}, e\textsubscript{(2, 4)}, e\textsubscript{(2, 8)}\} are included into the final EDD solution. When the algorithm reaches \textit{node 3}, it encounters two nodes, i.e., \textit{node 5} and \textit{node 9} exceeding the hop-limit. Now, since \textit{node 5} is encountered first, the algorithm adds \textit{node 5} to the cloud directly, which then updates the final EDD solution to include \textit{node 6} and \textit{node 9}, as they can be reached within vendors' latency constraint with \textit{node 5} as the root. The final EDD solution then becomes  \{e\textsubscript{(c, 2)}, e\textsubscript{(c, 5)}, e\textsubscript{(2, 3)}, e\textsubscript{(2, 4)}, e\textsubscript{(2, 8)}, e\textsubscript{(5, 6)}, e\textsubscript{(5, 9)}\}, which incurs a total cost of $2\gamma$ + $5$, including both the C2E and E2E edges.

\subsection{Time Complexity analysis of EDD-NSTE algorithm}

There has been a wealth of research in approximating the best Steiner Tree for a given general graph as mentioned in Gropl et al. \cite{IEEEexample:collectionpaperstudy}, with a lower bound of smaller than 1.01 \cite{IEEEexample:paerlowerbound} theoretically. In this paper, we first introduced an algorithm to calculate the Network Steiner Tree approximation, based on the algorithm proposed in Zelikovsky et al. \cite{IEEEexample:paperZelikovsky} which has a time complexity of $O(\norm{V}\norm{E} + R\textsuperscript{4})$ and an approximation ratio of $O(11/6)$. The same approximation ratio can be achieved with a faster implementation of $O(\norm{R}(\norm{E} + \norm{V}\norm{R} +  \norm{V}log\norm{V}))$ as mentioned in \cite{IEEEexample:paperZelikovskysecond}.

The algorithm EDD-NSTE has a time complexity of $O(\norm{V} + \norm{E})$ in the worst case. Thus, the total time complexity of the EDD-NSTE Algorithm along with network Steiner tree estimation is $O((\norm{V} + \norm{E}) + \norm{V}\norm{E} + \norm{R}\textsuperscript{4})$ and in the worst case the time complexity becomes $O(\norm{V}\textsuperscript{4})$. 

In the next subsection, we prove that the EDD-NSTE is an $O(k)$ approximation algorithm with experimental results suggesting that EDD-NSTE significantly outperforms the other three representative approaches in comparison with a performance margin of 86.67\% (as per our experimental result given Section V).

\subsection{EDD-NSTE is an O(k) approximation algorithm}

As we discussed in the previous section, let tree $G\textsubscript{final}$ be the final EDD solution produced by \hyperlink{algo:algo3}{Algorithm 3}, $G\textsubscript{ST}$ be the minimum cost steiner tree approximated by \hyperlink{algo:algo1}{Algorithm 1}, and $G\textsubscript{cv}$ be the tree having cloud $c$ as the root and edges $e\textsubscript{(c, v)}$ , $\forall$ v $\in$ $G\textsubscript{final} \setminus \{c\}$. Also, let the cost associated with each of these trees be $Cost(G\textsubscript{final})$, $Cost(G\textsubscript{ST})$, and $Cost(G\textsubscript{cv})$ respectively. 

Let $G\textsubscript{opt}$ be the optimal solution for the given EDD problem and the optimal cost associated be $Cost(G\textsubscript{opt})$, thus, as per the research in  Zelikovsky et al. \cite{IEEEexample:paperZelikovsky}, we can write
\begin{equation}
    Cost(G\textsubscript{ST}) = \frac{11}{6}\hspace{2pt}Cost(G\textsubscript{opt})
\end{equation}
Now, let $v\textsubscript{o} = c$ be the cloud server and $v\textsubscript{i}$ as the $i^{th}$ edge server, where $i \in \{1, 2, \cdots, m\}$, which brings extra paths during the DFS iteration. Also, let $K = D\textsubscript{limit}$ be the depth limit provided by the vendor. Thus, for any edge server there exists two possibilities, i.e.,
\begin{itemize}
    \item $v\textsubscript{i}$ exceeds the latency constraint of $D\textsubscript{limit}$ and thus, we need to add an extra edge $(c, v\textsubscript{i})$, which in turn adds a cost of $Cost\textsubscript{(G\textsubscript{cv})}(c, v\textsubscript{i}) = \gamma$.
    \item $\exists$ a path $(c, v\textsubscript{i})$ as a path from $(c, v\textsubscript{i-1})$ and  $(v\textsubscript{i-1}, v\textsubscript{i})$, such that $v\textsubscript{i}$ does not exceed the latency constraint, which in turn adds a cost of $Cost\textsubscript{(G\textsubscript{final})}(c, v\textsubscript{i}) \leq Cost\textsubscript{(G\textsubscript{cv})}(c, v\textsubscript{i-1}) + Cost\textsubscript{(G\textsubscript{ST})}(v\textsubscript{i-1}, v\textsubscript{i})$.
\end{itemize}

In the first case, when edge server $v\textsubscript{i}$ exceeds the latency constraint, then we must have
\begin{equation}
    Cost\textsubscript{(G\textsubscript{ST})}(c, v\textsubscript{i}) \geq \gamma + K \geq \left(1 + \frac{K}{\gamma}\right)Cost\textsubscript{(G\textsubscript{cv})}(c, v\textsubscript{i})
\end{equation}
where, $K$ denotes the latency constraint $D\textsubscript{limit}$. Thus, we can obtain the combined equation as
\begin{equation}
\begin{gathered}
    \left(1 + \frac{K}{\gamma}\right)Cost\textsubscript{(G\textsubscript{cv})}(c, v\textsubscript{i}) \leq \\ Cost\textsubscript{(G\textsubscript{ST})}(c, v\textsubscript{i}) \leq Cost\textsubscript{(G\textsubscript{cv})}(c, v\textsubscript{i-1}) + Cost\textsubscript{(G\textsubscript{ST})}(v\textsubscript{i-1}, v\textsubscript{i})
\end{gathered}
\end{equation}
Summing the above equation for all of the $m$ edge servers, i.e., $v\textsubscript{i}$, where $i \in \{1, 2, ...m\}$, we have
\begin{equation}
\begin{gathered}
    \left(1 + \frac{K}{\gamma}\right) \sum_{i=1}^{m}Cost\textsubscript{(G\textsubscript{cv})}(c, v\textsubscript{i}) \leq \\ \sum_{i=1}^{m}Cost\textsubscript{(G\textsubscript{cv})}(c, v\textsubscript{i-1}) + \sum_{i=1}^{m}Cost\textsubscript{(G\textsubscript{ST})}(v\textsubscript{i-1}, v\textsubscript{i})
\end{gathered}
\end{equation}
Now, as we know from the basic inequality of positive numbers, that $\sum_{i=1}^{m-1}Cost\textsubscript{(G\textsubscript{cv})}(c, v\textsubscript{i}) \leq \sum_{i=1}^{m}Cost\textsubscript{(G\textsubscript{cv})}(c, v\textsubscript{i})$, so the above equation yields \textemdash
\begin{equation}
    \frac{K}{\gamma}\sum_{i=1}^{m}Cost\textsubscript{(G\textsubscript{cv})}(c, v\textsubscript{i}) \leq \sum_{i=1}^{m}Cost\textsubscript{(G\textsubscript{ST})}(v\textsubscript{i-1}, v\textsubscript{i})
\end{equation}
\begin{equation}
    \frac{K}{\gamma}Cost(G\textsubscript{cv}) \leq \sum_{i=1}^{m}Cost\textsubscript{(G\textsubscript{ST})}(v\textsubscript{i-1}, v\textsubscript{i})
\end{equation}
For each edge server $v\textsubscript{i}$, that changes its path in $G\textsubscript{final}$ during the DFS traversal without adding the path $(c, v\textsubscript{i})$, the total cost of update is less than that of $Cost\textsubscript{(G\textsubscript{cv})}(c, v\textsubscript{i})$ and thus, the cost after forming $G\textsubscript{ST}$ must not be more than $\sum_{i=1}^{m}Cost\textsubscript{(G\textsubscript{cv})}(c, v\textsubscript{i})$. Now, each edge in $G\textsubscript{ST}$ is traversed at most twice during the DFS traversal, thus, we have
\begin{equation}
    \sum_{i=1}^{m}Cost\textsubscript{(G\textsubscript{ST})}(v\textsubscript{i-1}, v\textsubscript{i}) \leq 2 \cdot Cost(G\textsubscript{ST})    
\end{equation}
Thus, combining both above equations, results in 
\begin{equation}
    Cost(G\textsubscript{cv}) \leq \frac{2\gamma}{K} Cost(G\textsubscript{ST})
\end{equation}
Finally, as we know, that the cost of the final EDD solution approximated by EDD-NSTE is
\begin{equation}
    Cost(G\textsubscript{final}) \leq Cost(G\textsubscript{cv}) + Cost(G\textsubscript{ST})
\end{equation}
\begin{equation}
    Cost(G\textsubscript{final}) \leq \frac{2\gamma}{K} Cost(G\textsubscript{ST}) + Cost(G\textsubscript{ST})
\end{equation}
\begin{equation}
    Cost(G\textsubscript{final}) \leq \left(\frac{2\gamma}{K} + 1\right) Cost(G\textsubscript{ST})
\end{equation}
\begin{equation}
    Cost(G\textsubscript{final}) \leq \frac{11}{6} \left(\frac{2\gamma}{K} + 1\right) Cost(G\textsubscript{opt})
\end{equation}
Now, since $K$ and $\gamma$ are both constants, we can safely assume $k = \frac{11}{6} \left(\frac{2\gamma}{K} + 1\right)$, and thus, EDD-NSTE is an $O(k)$ approximation algorithm.

As compared to Xia et al. \cite{IEEEexample:papermy}, our approximation is lightly better as $\frac{11}{6} \left(\frac{2\gamma}{K} + 1\right) \leq 2 \left(\frac{2\gamma}{K} + 1\right) $. But in actual run this produce even better result as compared to EDD-A \cite{IEEEexample:papermy}.   

\section{Experimental Evaluation}
\label{sec:EE}

We conducted experiments to evaluate the effectiveness of the proposed EDD-NSTE approach in comparisons with the  state of art EDD-A \cite{IEEEexample:papermy}, refined LP solution and other representative approaches. All these experiments were conducted on an Ubuntu 20.04.1 LTS machine equipped with Intel Core i5-7200U processor (4 CPU’s, 2.50GHz) and 16GB RAM. The integer programming simulation was done on Google Collab (Python 3 Google Computer Engine Backend, 12.72GB RAM).

\subsection{Simulation Settings and Approaches for comparison of the result}

Description of the state of art solution technique (EDD-A) and other representative approaches are \textemdash
\begin{itemize}
    \item \textbf{Greedy Connectivity (GC)} \textemdash \hspace{2pt} In this approach, for each node, the approach define its connectivity which is equal to the number of destination servers that have not received the data yet and are reachable within the vendors' latency constraint. Then it select the node with the maximum connectivity and make it an initial transmit edge server which then transmits the data to other servers within the app vendors' latency limit of \textit{d\textsubscript{limit}} until all destination edge servers have received the data.
    \item \textbf{Random} \textemdash \hspace{2pt} In this approach, it randomly select the initial transit edge servers which directly receive data from the cloud and then transmit it to the other servers within the app vendors' latency limit of \textit{d\textsubscript{limit}}, until all destination edge servers have received the data.
    \item \textbf{EDD Integer Programming} \textemdash \hspace{2pt} As discussed above in the \hyperlink{SolSgy}{Section IV}, we refine the formulation of given EDD problem into a 0-1 integer programming problem which can then be solved by using any simple IP solvers. This method take a huge amount of time to produce the result in the case of large datasets as is exponential in complexity. 
    \item \textbf{EDD-A Algorithm} \textemdash \hspace{2pt} This is state of the art approach given in \cite{IEEEexample:papermy}, in this approach it first calculate a connectivity-oriented minimum Steiner tree (CMST) using an $O(2)$ approximation method. This method calculates a minimum Steiner tree having cost at most twice the optimal Steiner tree on the graph. It then uses a heuristic algorithm to calculate the minimum cost of transmission incurred using E2E and C2E edges in the CMST. This algorithm is proved to have an $O(k)$ approximation solution and a time complexity $O(\norm{V}\textsuperscript{3})$, in the worst case. 
\end{itemize}

\begin{figure*}[t]
\centering
\begin{minipage}[b]{0.48\textwidth}
\centering
\begin{tikzpicture}
        \begin{axis}[
            xlabel={Number of nodes $N$ in the graph $G$},
            ylabel={Total EDD cost incurred (times 1-Hop E2E)},
            xmin=0, xmax=1000,
            ymin=0, ymax=400,
            xtick={},
            ytick={},
            width=\textwidth,
            height=0.9\textwidth,
            legend pos=south east,
            ymajorgrids=true,
            grid style=dashed,
            mark size = 1.5pt,
            every axis plot/.append style={line width=0.5pt},
        ]
        \addplot[
            color=ip,
            every mark/.append style={solid, mark size = 2pt, fill=ip}, mark=triangle*
            ]
            coordinates {
            (0,0)(10,45)(100,130)(500,189)(1000,217)
            };
        \addplot[
            color=eddste,
            every mark/.append style={solid, mark size = 2pt, fill=eddste}, mark=diamond*,
            ]
            coordinates {
            (0,0)(10,45)(100,167)(500,232)(1000,270)
            };
        \addplot[
            color=eddA,
            every mark/.append style={solid, fill=eddA}, mark=square*,
            ]
            coordinates {
            (0,0)(10,45)(100,203)(500,261)(1000,295)
            };
        \addplot[
            color=greedy,
            every mark/.append style={solid, fill=greedy}, mark=otimes*,
            ]
            coordinates {
            (0,0)(10,46)(100,249)(500,445)(1000,1644)
            };
        \addplot[
            color=random,
            every mark/.append style={solid, mark size = 2.5pt, fill=random}, mark=x,
            ]
            coordinates {
            (0,0)(10,91)(100,397)(500,989)(1000,4556)
            };
        \legend{Integer Programming, EDD-NSTE Algorithm, EDD-A Algorithm, Greedy Connectivity, Random}
        \end{axis}
        \end{tikzpicture}
\caption{$N$ v/s Total EDD cost}\label{fig:fig5}
\end{minipage}\qquad
\hfill
\begin{minipage}[b]{0.48\textwidth}
\centering
\begin{tikzpicture}
        \begin{axis}[
            xlabel={ Number of destination edge servers $R$},
            ylabel={Total EDD cost incurred (times 1-Hop E2E)},
            xmin=0, xmax=50,
            ymin=0, ymax=800,
            xtick={},
            ytick={},
            width=\textwidth,
            height=0.9\textwidth,
            legend pos=north west,
            ymajorgrids=true,
            grid style=dashed,
            mark size = 1.5pt,
            every axis plot/.append style={line width=0.5pt},
        ]
        \addplot[
            color=ip,
            every mark/.append style={solid, mark size = 2pt, fill=ip}, mark=triangle*,
            ]
            coordinates {
            (0,0)(5,46)(10,90)(25,130)(30,133)(50,191)
            };
        \addplot[
            color=eddste,
            every mark/.append style={solid, mark size = 2pt, fill=eddste}, mark=diamond*,
            ]
            coordinates {
            (0,0)(5,81)(10,108)(25,167)(30,247)(50,327)
            };
        \addplot[
            color=eddA,
            every mark/.append style={solid, fill=eddA}, mark=square*,
            ]
            coordinates {
            (0,0)(5,81)(10,108)(25,203)(30,227)(50,380)
            };
        \addplot[
            color=greedy,
            every mark/.append style={solid, fill=greedy}, mark=otimes*,
            ]
            coordinates {
            (0,0)(5,86)(10,159)(25,249)(30,218)(50,350)
            };
        \addplot[
            color=random,
            every mark/.append style={solid, mark size=2.5pt, fill=random}, mark=x,
            ]
            coordinates {
            (0,0)(5,234)(10,317)(25,434)(50,513)
            };
        \legend{Integer Programming, EDD-NSTE Algorithm, EDD-A Algorithm, Greedy Connectivity, Random}
        \end{axis}
        \end{tikzpicture}
\caption{$R$ v/s Total EDD cost}\label{fig:fig6}
\end{minipage}
\begin{minipage}[b]{\textwidth}
\centering
\begin{tikzpicture}
        \begin{axis}[
            xlabel={Latency constraint $D\textsubscript{limit}$},
            ylabel={Total EDD cost incurred (times 1-Hop E2E)},
            xmin=0, xmax=10,
            ymin=0, ymax=800,
            xtick={},
            ytick={},
            width=\textwidth,
            height=0.4\textwidth,
            legend pos=north east,
            ymajorgrids=true,
            grid style=dashed,
            mark size = 1.5pt,
            every axis plot/.append style={line width=0.5pt},
        ]
        \addplot[
            color=ip,
            every mark/.append style={solid, mark size = 2pt, fill=ip}, mark=triangle*,
            ]
            coordinates {
            (1,500)(2,240)(3,130)(4,92)(5,90)(6,67)(7,67)(8,62)(9,59)(10,52)
            };
        \addplot[
            color=eddste,
            every mark/.append style={solid, mark size = 2pt, fill=eddste}, mark=diamond*,
            ]
            coordinates {
            (1,500)(2,386)(3,167)(4,111)(5,97)(6,72)(7,72)(8,79)(9,79)(10,75)
            };
        \addplot[
            color=eddA,
            every mark/.append style={solid, fill=eddA}, mark=square*,
            ]
            coordinates {
            (1,500)(2,386)(3,203)(4,209)(5,150)(6,125)(7,100)(8,107)(9,79)(10,63)
            };
        \addplot[
            color=greedy,
            every mark/.append style={solid, fill=greedy}, mark=otimes*,
            ]
            coordinates {
            (1,500)(2,318)(3,249)(4,215)(5,205)(6,217)(7,217)(8,217)(9,217)(10,217)
            };
        \addplot[
            color=random,
            every mark/.append style={solid, mark size = 2.5pt, fill=random}, mark=x,
            ]
            coordinates {
            (1,765)(2,749)(3,397)(4,344)(5,321)(6,303)(7,217)(8,217)(9,217)(10,217)
            };
        \legend{Integer Programming, EDD-NSTE Algorithm, EDD-A Algorithm, Greedy Connectivity, Random}
        \end{axis}
        \end{tikzpicture}
\caption{Latency constraint $D\textsubscript{limit}$ v/s Total EDD cost, lower is better}\label{fig:fig7}
\vspace{0.2cm}
\end{minipage}
\end{figure*}

\subsection{Dataset Used}

The experiments were performed on a standard real-world EUA dataset \footnote{https://github.com/swinedge/eua-dataset}, which contains the geo-locations of more than 1,400 edge-server base stations of Melbourne, Australia. The set of destination edge servers \textit{R} is generated randomly. The edges between the nodes of the graph are also generated randomly using an online random connected graph model generator tool. In the experiments, the value of  $\gamma$ is set to 20 which is same setting as specified in \cite{IEEEexample:papermy}.

\begin{figure*}[tb!]
\centering
\begin{minipage}[b]{0.48\textwidth}
\centering
\begin{tikzpicture}
    \begin{axis}[
    	symbolic x coords={0.05, 0.10, 0.25, 0.30, 0.50, 0.60},
    	ylabel=Total EDD cost (times 1-Hop E2E),
    	enlargelimits=0.05,
    	legend style={at={(0.5,-0.2)},
    	anchor=north,legend columns=-1},
    	ybar interval=0.7,
    	width=\textwidth,
    	height=0.6\textwidth,
    ]
    \addplot 
    	coordinates {(0.05,46) (0.10,90)
    		 (0.25,130) (0.30,133) (0.50,191) (0.60, 700)};
    \addplot 
    	coordinates {(0.05,81) (0.10,108)
    		 (0.25,167) (0.30,150) (0.50,327)(0.60, 700)};
    \addplot 
    	coordinates {(0.05,81) (0.10,108)
    		 (0.25,203) (0.30,227) (0.50,350)(0.60, 700)};
    \addplot 
    	coordinates {(0.05,86) (0.10,159)
    		 (0.25,249) (0.30,218) (0.50,350)(0.60, 700)};
    \addplot 
    	coordinates {(0.05,234) (0.10,367)
    		 (0.25,494) (0.30,553) (0.50,759)(0.60, 700)};
    		 
    \legend{IP, EDD-NSTE, EDD-A, GC, Random}
    \end{axis}
\end{tikzpicture}
\caption{$\rho$ v/s Total EDD cost}\label{fig:fig8}
\end{minipage}\qquad
\hfill
\begin{minipage}[b]{0.48\textwidth}
\centering
\begin{tikzpicture}
    \begin{axis}[
    	symbolic x coords={1.00, 1.50, 2.00, 2.50, 3.00, 4.00},
    	ylabel=Total EDD cost (times 1-Hop E2E),
    	enlargelimits=0.05,
    	legend style={at={(0.5,-0.2)},
    	anchor=north,legend columns=-1},
    	ybar interval=0.7,
    	width=\textwidth,
    	height=0.6\textwidth,
    ]
    \addplot
    	coordinates {(1.00,959) (1.50,653)
    		 (2.00,594) (2.50,467) (3.00,334)(4.00, 700)};
    \addplot 
    	coordinates {(1.00,650) (1.50,418)
    		 (2.00,349) (2.50,259) (3.00,210)(4.00, 700)};
    \addplot 
    	coordinates {(1.00,550) (1.50,427)
    		 (2.00,303) (2.50,208) (3.00,189)(4.00, 700)};
    \addplot 
    	coordinates {(1.00,527) (1.50,447)
    		 (2.00,267) (2.50,200) (3.00,181)(4.00, 700)};
    
    \addplot 
    	coordinates {(1.00,491) (1.50,353)
    		 (2.00,230) (2.50,190) (3.00,146) (4.00, 700)};
    		 
    \legend{Random,GC, EDD-A, EDD-NSTE, IP}
    \end{axis}
\end{tikzpicture}
\caption{$\delta$ v/s Total EDD cost, lower is better}\label{fig:fig9}
\end{minipage}
\end{figure*}

\begin{figure*}[tb!]
\centering
\begin{minipage}[b]{0.22\textwidth}
\centering
\begin{tikzpicture}
    \begin{axis}[
            xlabel={$N$},
            ylabel={Time (in s)},
            xmin=0, xmax=100,
            ymin=0, ymax=10,
            xtick={0, 50, 100},
            ytick={},
            width=\textwidth,
            height=\textwidth,
            legend pos=north west,
            ymajorgrids=true,
            grid style=dashed,
            mark size = 1pt,
            every axis plot/.append style={line width=0.5pt},
        ]
        \addplot[
            color=eddste,
            every mark/.append style={solid, fill=eddste}, mark=square*,
            ]
            coordinates {
            (0,0)(10,0.5)(25,0.60)(50,1)(70,1.9)(85,5)(100,10)
            };
        \end{axis}
\end{tikzpicture}
\caption{$N$ v/s Computational Overhead, lower is better}\label{fig:fig10}
\end{minipage}\qquad
\hfill
\begin{minipage}[b]{0.22\textwidth}
\centering
\begin{tikzpicture}
    \begin{axis}[
            xlabel={$\rho$},
            ylabel={Time (in s)},
            xmin=0, xmax=2.0,
            ymin=0, ymax=10,
            xtick={0.0, 0.5, 1.0, 1.5, 2.0},
            ytick={},
            width=\textwidth,
            height=\textwidth,
            legend pos=north west,
            ymajorgrids=true,
            grid style=dashed,
            mark size = 1pt,
            every axis plot/.append style={line width=0.5pt},
        ]
        \addplot[
            color=eddste,
            every mark/.append style={solid, fill=eddste}, mark=square*,
            ]
            coordinates {
            (0,0)(0.5,0.6)(1.0,1.5)(1.5,2.9)(1.8,4.6)(2.0,8.9)
            };
        \end{axis}
\end{tikzpicture}
\caption{$\rho$ v/s Computational Overhead,lower is better}\label{fig:fig11}
\end{minipage}\qquad
\hfill
\begin{minipage}[b]{0.22\textwidth}
\centering
\begin{tikzpicture}
    \begin{axis}[
            xlabel={$D\textsubscript{limit}$},
            ylabel={Time (in s)},
            xmin=0, xmax=5,
            ymin=0, ymax=10,
            xtick={1, 2, 3, 4, 5},
            ytick={},
            width=\textwidth,
            height=\textwidth,
            legend pos=north west,
            ymajorgrids=true,
            grid style=dashed,
            mark size = 1pt,
            every axis plot/.append style={line width=0.5pt},
        ]
        \addplot[
            color=eddste,
            every mark/.append style={solid, fill=eddste}, mark=square*,
            ]
            coordinates {
            (0,0)(1,1.8)(2,2.6)(3,6.2)(4,5.4)(5,3.4)
            };
        \end{axis}
\end{tikzpicture}
\caption{$D\textsubscript{limit}$ v/s Computational Overhead, lower is better}\label{fig:fig12}
\end{minipage}\qquad
\hfill
\begin{minipage}[b]{0.22\textwidth}
\centering
\begin{tikzpicture}
    \begin{axis}[
            xlabel={$\delta$},
            ylabel={Time (in s)},
            xmin=0, xmax=2.0,
            ymin=0, ymax=100,
            xtick={0.0, 0.5, 1.0, 1.5, 2.0},
            ytick={},
            width=\textwidth,
            height=\textwidth,
            legend pos=north west,
            ymajorgrids=true,
            grid style=dashed,
            mark size = 1pt,
            every axis plot/.append style={line width=0.5pt},
        ]
        \addplot[
            color=eddste,
            every mark/.append style={solid, fill=eddste}, mark=square*,
            ]
            coordinates {
            (0,0)(0.5,20)(1.0,30)(1.5,50)(2.0,100)
            };
        \end{axis}
\end{tikzpicture}
\caption{$\delta$ v/s Computational Overhead, lower is better}\label{fig:fig13}
\end{minipage}
\vspace{-0.8cm}
\end{figure*}

\subsection{Experimental Results}

Different EDD scenarios and algorithms mentioned above in the section \hyperlink{SolSgy}{solution strategy} and other approaches are simulated and compared to each other by varying different parameters as mentioned below \textemdash
\begin{itemize}
    \item The number of nodes $N$ in the graph $G$.
    \item The number of destination edge servers $R$ in the graph.
    \item The vendor's latency constraint limit, i.e., $D\textsubscript{limit}$.
\end{itemize}

Among the above three factors, we vary one of the factors while keeping the others constant and then compare the estimated total EDD cost incurred in each case with the optimal cost generated by the integer programming approach.

\autoref{fig:fig5}, displays the relationship between the number of nodes $N$ in the graph $G$ v/s total EDD cost incurred by different simulation settings and approaches. The value of the other parameters such as $D\textsubscript{limit} = 3$, $R = 25$ random nodes, and $\gamma = 20$ were fixed during the experiment. For $N < 100$, there is a very slight difference between other approaches but as the value of $N$ increases  the difference becomes significantly more. Resultant cost produced by EDD-NSTE approach is lower (is better) as compared to EDD-A in all the cases.

Similarly, \autoref{fig:fig6}, displays the relationship between the number of destination edge servers $R$ v/s the total EDD cost incurred by different other settings. For this experiment the value of other parameters such as $N = 100$, $D\textsubscript{limit} = 3$ and $\gamma = 20$ were fixed during the experiment. For $\norm{R} \leq 10$, the EDD cost incurred in all approaches except the Random approach results in the same cost but the number of destination edge servers increase in the graph the difference becomes quite broad. The Greedy Connectivity algorithm for the given dataset with $\norm{R} = 30$, performs better than other approximation algorithms like EDD-A and EDD-NSTE. This may be an anomaly.  However, the Greedy Algorithm, cannot guarantee to maintain a certain performance ratio on any given dataset, unlike EDD-A or EDD-NSTE.  Resultant cost produced by EDD-NSTE approach is lower (is better) as compared to EDD-A in most of the cases except $\norm{R} = 30$.

Now, the only factor left to alter is $D\textsubscript{limit}$, and the rest of the parameters are kept constant similar to the previous case. The number of destination edge servers $\norm{R}$ $= 25$ for this experiment. \autoref{fig:fig7}, displays the relationship between the vendor's latency constraint $D\textsubscript{limit}$ v/s the total EDD cost incurred by different solution approaches. As we know, these two quantities are inversely related, i.e., with an increase in the value of latency constraint, the total EDD cost decreases. The total EDD cost reaches its maxima of $\gamma$R when $D\textsubscript{limit}$ = 1, as in this case all the destination edge servers are connected directly to the cloud server to fulfill the vendors' latency constraint. Also,  Random approach and Greedy Connectivity algorithms become constant for $N \geq 7$, as then almost both the Random or Greedy selection of servers is not able to reduce the EDD cost.  Resultant cost produced by EDD-NSTE approach is lower (is better) as compared to EDD-A in all the value of $D\textsubscript{limit}$ from 2 to 9.

Here, we consider another parameter referred to as the destination edge server density. It is defined as the ratio of number of destination edge server upon total number of edge servers, i.e., Destination Edge Server Density ($\rho$) \textemdash
\begin{equation}
    \rho = \frac{\norm{R}}{\norm{V}}
\end{equation}

\autoref{fig:fig8}, displays the bar-graph relationship between the total EDD cost and the density of the destination edge servers. With increase in the density, the total EDD cost associated with each of the approach tend to increase as a whole. Among the other algorithms and approaches in comparison, EDD-NSTE Algorithm performed well overall. As $\rho$ increases the resultant cost difference between EDD-NSTE approach and LP approach is lower as compared to  the resultant cost difference between EDD-A approach and LP approach.

Similarly, we consider another experimental parameter known as edge density. This parameter helps to estimate the overall density of the graph. A very dense graph will have a higher value of $\delta$ than a sparse graph or a tree. Thus, it is defined as the total no of edges or high-speed links in the graph upon total no of edge servers, i.e, Edge Density ($\delta$) \textemdash
\begin{equation}
    \delta = \frac{\norm{E}}{\norm{V}}
\end{equation}

\autoref{fig:fig9}, displays the bar-graph relationship between the total EDD cost incurred and the density of the graph, or edge density. With the increase in the edge density, the total EDD cost associated with each of the simulations tends to decrease as now shorter paths from cloud to other edge servers exist and thus, any destination edge server can be reached within the vendors' latency constraint following these shorter paths, thereby, decreasing the number of initial transit edge servers, which in turn decreases the total EDD cost incurred as the $Cost\textsubscript{C2E}$ is significantly greater than $Cost\textsubscript{E2E}$.  Here also, we see as $\rho$ increases the resultant cost difference between EDD-NSTE approach and LP approach is lower as compared to  the resultant cost difference between EDD-A approach and LP approach.

\autoref{fig:fig10} to \autoref{fig:fig13} displays the computational overhead of EDD-NSTE algorithm versus different other graph parameters like number of nodes($N$), destination edge server density($\rho$), the vendors' latency constraint($D\textsubscript{limit}$), and edge density($\delta$) respectively. As we can see, with an increase in $N$, $\delta$ or $\rho$, the computational overhead of EDD-NSTE increases, as its worst-case time complexity is $O(\norm{V}\textsuperscript{4})$, whereas, with an increase in $D\textsubscript{limit}$, the computational overhead first increases and then decreases, as relaxing the vendors' latency constraint after a certain limit decreases the extra overhead caused by the slicing and pruning algorithm.

\section{Conclusion}
\label{sec:CS}

The EDD-NSTE algorithm suggested in this paper performs considerably better than the other simulations and approaches implemented for the edge data distribution problem. The total EDD cost approximated by this algorithm is closer to the optimal solution than the basic hard-coded approaches like Greedy Connectivity or Random. The algorithm also performs significantly better than the last best EDD-A algorithm suggested in \cite{IEEEexample:papermy}. The EDD-A algorithm is an $O(k)$ approximation algorithm\footnote{$O(k)$ approximation algorithm means that solution produced by the algorithm is atmost $k$ times the optimal solution.} and the EDD-NSTE algorithm is an $O(k')$ approximation algorithm where $k' < k$ and thus, the performance of EDD-NSTE is better than EDD-A, which coincides with the experimental evaluations done so far. 

EDD-NSTE algorithm produced better solutions than EDD-A or other approaches in more than 86.67\% of the test cases, ranging from very dense graphs to sparse graphs or trees.

Greedy Connectivity may produce solutions better than EDD-A or EDD-NSTE in some of the cases, but the greedy solutions are just an ad-hoc solution and thereby, cannot always guarantee to produce results that closer to the actual solution whereas the latter algorithms are restricted by their approximation ratio to always produce solutions below a certain maximum value for a given graph, thus they perform better than these methods on an average.

\bibliographystyle{./bibtex/bib/IEEEtran}
\setstretch{0.8}
\bibliography{./bibtex/bib/IEEEabrv,./bibtex/bib/IEEEexample}

\begin{thebibliography}{10}
\providecommand{\url}[1]{#1}
\csname url@samestyle\endcsname
\providecommand{\newblock}{\relax}
\providecommand{\bibinfo}[2]{#2}
\providecommand{\BIBentrySTDinterwordspacing}{\spaceskip=0pt\relax}
\providecommand{\BIBentryALTinterwordstretchfactor}{4}
\providecommand{\BIBentryALTinterwordspacing}{\spaceskip=\fontdimen2\font plus
\BIBentryALTinterwordstretchfactor\fontdimen3\font minus
  \fontdimen4\font\relax}
\providecommand{\BIBforeignlanguage}[2]{{%
\expandafter\ifx\csname l@#1\endcsname\relax
\typeout{** WARNING: IEEEtran.bst: No hyphenation pattern has been}%
\typeout{** loaded for the language `#1'. Using the pattern for}%
\typeout{** the default language instead.}%
\else
\language=\csname l@#1\endcsname
\fi
#2}}
\providecommand{\BIBdecl}{\relax}
\BIBdecl

\bibitem{IEEE-CC-ACM}
M.~Armbrust and et~al., ``{A view of cloud computing},'' \emph{Commun. ACM},
  vol.~53, no.~4, pp. 50--58, 2010.

\bibitem{amazonec2}
\BIBentryALTinterwordspacing
Amazon ec2. [Online]. Available: \url{https://aws.amazon.com/ec2/}
\BIBentrySTDinterwordspacing

\bibitem{rfc8793}
B.~Wissingh, C.~A. Wood, and et~al., ``Information-centric networking {(ICN):}
  content-centric networking (ccnx) and named data networking {(NDN)}
  terminology,'' \emph{{RFC}}, vol. 8793, pp. 1--17, 2020.

\bibitem{Pathan2008}
M.~Pathan and R.~Buyya, \emph{{A Taxonomy of CDNs}}.\hskip 1em plus 0.5em minus
  0.4em\relax Berlin, Heidelberg: Springer Berlin Heidelberg, 2008, pp. 33--77.

\bibitem{IEEEexample:davisEC}
A.~Davis, J.~Parikh, and W.~E. Weihl, ``{Edge Computing: Extending Enterprise
  Applications to the Edge of the Internet},'' ser. WWW Alt. '04, 2004, p.
  180–187.

\bibitem{IEEE-EC}
W.~{Shi}, J.~{Cao}, Q.~{Zhang}, Y.~{Li}, and L.~{Xu}, ``{Edge Computing: Vision
  and Challenges},'' \emph{IEEE Internet of Things Journal}, vol.~3, no.~5, pp.
  637--646, 2016.

\bibitem{FH}
``{https://www.oculus.com/facebook-horizon/}.''

\bibitem{IEEEexample:papermy}
H.~Xia, F.~Chen, Q.~He, J.~C. Grundy, M.~Abdelrazek, and H.~Jin,
  ``{Cost-Effective App Data Distribution in Edge Computing},'' \emph{IEEE
  Tran. on Parallel and Distributed Systems}, vol.~32, no.~1, pp. 31--43, 2021.

\bibitem{IEEEexample:paperpiyush}
D.~Zhao, M.~Mohamed, and H.~Ludwig, ``Locality-aware scheduling for containers
  in cloud computing,'' \emph{IEEE Transactions on Cloud Computing}, vol.~8,
  no.~2, pp. 635--646, 2018.

\bibitem{IEEEexample:paperA}
H.~Yao, C.~Bai, M.~Xiong, D.~Zeng, and Z.~Fu, ``Heterogeneous cloudlet
  deployment and user-cloudlet association toward cost effective fog
  computing,'' \emph{Concurrency Comp. Pract. Exp.}, vol.~29, no.~16, 2017,.

\bibitem{IEEEexample:paperB}
H.~Yin, X.~Zhang, H.~H. Liu, Y.~Luo, C.~Tian, S.~Zhao, and F.~Li, ``Edge
  provisioning with flexible server placement,'' \emph{IEEE Trans. on Parallel
  Distribution System}, vol.~28, no.~4, p. 1031–1045, 2017.

\bibitem{IEEEexample:paperF}
P.~Lai, Q.~He, M.~Abdelrazek, F.~Chen, and et~al., ``Optimal edge user
  allocation in edge computing with variable sized vector bin packing,'' in
  \emph{Proc. Int. Conf. Service-Oriented Comput.}, pp. 230-245, 2018, Dec.
  5--9, 2018, pp. 230--245.

\bibitem{IEEEexample:papersparsh}
P.~Lai, Q.~He, J.~Grundy, and et~al., ``Cost-effective app user allocation in
  an edge computing environment,'' \emph{IEEE Transactions on Cloud Computing},
  vol.~31, no.~15, pp. 1--13, 2020.

\bibitem{IEEEexample:paperC}
X.~Cao, J.~Zhang, and H.~V. Poor, ``An optimal auction mechanism for mobile
  edge caching,'' in \emph{Proc. 38th IEEE Int. Conf. Distrib. Comput. Syst.},
  pp. 388-399, 2018, Dec. 5--9, 2018, pp. 388--399.

\bibitem{IEEEexample:paperD}
U.~Drolia, K.~Guo, J.~Tan, R.~Gandhi, and P.~Narasimhan, ``{Cachier:
  Edge-caching for recognition applications},'' in \emph{Proc. 37th IEEE Int.
  Conf. Distrib. Comput. Syst.}, pp. 276-286, 2017, Dec. 5--9, 2017, p.
  276–286.

\bibitem{IEEEExample:paperX}
K.~Zhang, S.~Leng, Y.~He, S.~Maharjan, and Y.~Zhang, ``{Cooperative Content
  Caching in 5G Networks with Mobile Edge Computing},'' \emph{IEEE Wireless
  Communications}, vol.~25, no.~3, pp. 80--87, 2018.

\bibitem{IEEEExample:paperY}
R.~Halalai, P.~Felber, A.-M. Kermarrec, and F.~Taïani, ``Agar: A caching
  system for erasure-coded data,'' in \emph{Proc. 37th IEEE Int. Conf. Distrib.
  Comput. Syst.}, pp. 23-33, 2017, Dec. 5--9, 2017, pp. 23--33.

\bibitem{IEEEExample:paperZ}
X.~Zhang and Q.~Zhu, ``{Hierarchical caching for statistical QoS guaranteed
  multimedia transmissions over 5G edge computing mobile wireless networks},''
  \emph{IEEE Wireless Comm.}, vol.~25, no.~3, pp. 12--20, 2018.

\bibitem{IEEEexample:paperE}
M.~Breitbach, D.~Schäfer, J.~Edinger, and C.~Becker, ``Context-aware data and
  task placement in edge computing environments,'' in \emph{Proc. IEEE Int.
  Conf. Pervasive Comput. Commun.}, pp. 1-10, 2019, Dec. 5--9, 2019, pp. 1--10.

\bibitem{IEEEexample:paperZelikovsky}
A.~Z. Zelikovsky, ``An 11/6-approximation algorithm for the network steiner
  problem,'' \emph{Algorithmica}, vol.~9, no.~5, pp. 463--470, 1993.

\bibitem{IEEEexample:collectionpaperstudy}
C.~Gröpl, S.~Hougardy, T.~Nierhoff, and J.~Prömel, ``Approximation algorithms
  for the steiner tree problems in graphs,'' \emph{Springer Verlag Berlin
  Heidelberg}, vol.~46, no.~5, pp. 235--279, 2001.

\bibitem{IEEEexample:paerlowerbound}
C.~Gröpl, S.~Hougardy, T.~Nierhoff, and H.~J. Prömel, ``Lower bounds for
  approximation algorithms for the steiner tree problem,'' \emph{Springer
  Verlag Berlin Heidelberg}, vol.~46, no.~5, pp. 217--228, 2001.

\bibitem{IEEEexample:paperZelikovskysecond}
A.~Z. Zelikovsky, ``A faster approximation algorithm for the steiner tree
  problem in graphs,'' \emph{Information Processing Letters}, vol.~46, no.~5,
  pp. 79--83, 1993.

\end{thebibliography}

\end{document}